\documentclass[11pt,notitlepage]{article}
\usepackage{amsmath}
\usepackage{bm}
\usepackage{tikz}
\usepackage{ulem}
\usetikzlibrary{calc}
\usepackage{authblk}
\usepackage{geometry}
\usepackage{stata}
\usepackage{verbatim}

\usetikzlibrary{arrows,positioning} 
	\tikzset{
	   	 >=stealth',
	    	 diskret/.style={rectangle,rounded corners,draw=black,text width=8em,minimum height=2em,text centered},
		 osynlig/.style={rectangle,rounded corners,draw=white,text width=8em,minimum height=2em,text centered},
		 streckad/.style={rectangle,rounded corners,draw=black,dashed,text width=8em,minimum height=2em,text centered},
	   	 stor/.style={rectangle, rounded corners,draw=white, text width=12em, minimum height=1em},
	   	 pil/.style={->,shorten <=3pt,shorten >=3pt,},
	   	 pils/.style={->,draw=aqua, thick,shorten <=3pt,shorten >=3pt,}
		}

\begin{document}
\title{A multi-state model incorporating estimation of excess hazards and multiple time scales}

\author[1,2]{Caroline E. Weibull*}
\author[1,3]{Paul C. Lambert}
\author[2]{Sandra Eloranta}
\author[1]{Therese M.L. Andersson}
\author[1]{Paul W. Dickman}
\author[1,3]{Michael J. Crowther}

\affil[1]{Department of Medical Epidemiology and Biostatistics, Karolinska Institutet, Box 281, 171 77 Stockholm, Sweden}
\affil[2]{Department of Medicine Solna, Clinical Epidemiology Division, Karolinska Institutet, Eugeniahemmet, T2, Karolinska Universitetssjukhuset, 171 76 Solna, Sweden}
\affil[3]{Biostatistics Research Group, Department of Health Sciences, University of Leicester, University Road, Leicester, LE1 7RH, UK}
\date{}
\maketitle

\begin{abstract}
As cancer patient survival improves, late effects from treatment are becoming the next clinical challenge. Chemotherapy and radiotherapy, for example, potentially increase the risk of both morbidity and mortality from second malignancies and cardiovascular disease. To provide clinically relevant population-level measures of late effects, it is of importance to (1) simultaneously estimate the risks of both morbidity and mortality, (2) partition these risks into the component expected in the absence of cancer and the component due to the cancer and its treatment, and (3) incorporate the multiple time scales of attained age, calendar time, and time since diagnosis. Multi-state models provide a framework for simultaneously studying morbidity and mortality, but do not solve the problem of partitioning the risks. However, this partitioning can be achieved by applying a relative survival framework, by allowing is to directly quantify the \textit{excess} risk. This paper proposes a combination of these two frameworks, providing one approach to address (1)-(3). Using recently developed methods in multi-state modeling, we incorporate estimation of excess hazards into a multi-state model. Both intermediate and absorbing state risks can be partitioned and different transitions are allowed to have different and/or multiple time scales. We illustrate our approach using data on Hodgkin lymphoma patients and excess risk of diseases of the circulatory system, and provide user-friendly Stata software with accompanying example code.
\end{abstract}


\section{Introduction}\label{intro}
Marked improvement in cancer patient survival in recent decades has increased the importance of studying treatment side-effects\cite{Aziz2003}, i.e., late effects. For example, chemotherapy and radiotherapy increase the risk of both secondary malignancies and cardiovascular disease (CVD)\cite{Floyd2005,Lee2013}. As therapies are developed and adapted to maximise the chances of cure while reducing risk of treatment toxicity, it is of interest to simultaneously estimate real-world morbidity and mortality. Today, a major challenge in the field of late effects lies in estimating the proportion of the observed morbidity and mortality that can be ascribed to the underlying cancer and its treatment. Hence, the standard approach is to simply estimate the cumulative \textit{observed} incidence and mortality\cite{vanNimwegen2015}. This article proposes a population-based approach to this problem bringing together statistical methods for relative survival and multi-state modelling. The former allows estimation of excess incidence, which under certain assumptions corresponds to the incidence above and beyond that expected in the absence of cancer and treatment thereof, and the latter enables investigation of patient trajectories going from an initial state (diagnosis), potentially via intermediate states (e.g., non-fatal illnesses such as CVD), to final absorbing states (different causes of death). Our focus is on the illness-death model where the illness state can be experienced also in the absence of cancer (unlike recurrence, which only occur in cancer patients). In addition, we know only when patients enter each state and not whether it was due to the cancer. We are interested in partitioning the probability of being in the illness state into that due to the cancer and its treatment, and that which would be expected in the absence of cancer.
 
Previous attempts to combine relative survival with multi-state modelling has proposed a Markov piecewise constant rate model incorporating estimation of excess mortality\cite{Huszti2012,Gilard-Pioc2015}. This method has been further generalised by Gillaizeau et al. to include semi-Markov models\cite{Gillaizeau2017}. In these papers, the relative survival methods are applied to absorbing states alone, i.e., excess mortality can be estimated but not excess incidence. Moreover, focus has been on relative effects rather than predictions. Our suggested approach enables partitioning of both excess morbidity and mortality in a Markov or semi-Markov setting, and we are unaware of existing methods for doing this. 

In population-based cancer patient survival, relative survival is a commonly used summary measure of patient survival as it captures both the direct and indirect contribution of a cancer diagnosis on mortality by comparing the observed survival of the patients to the expected survival in a comparable cancer-free population\cite{Sarfati2010,Dickman2004,Dickman2006}. In the standard setting, the outcome is all-cause death and the mortality rates expected in the absence of cancer are typically retrieved from publicly available population mortality tables. However, the same approach can be used to estimate excess incidence rates\cite{Weibull2019}. As such, the excess incidence of some disease of interest can be interpreted as the portion of incidence attributable to the underlying cancer and its treatment, given that the expected rates are appropriate for the population diagnosed with cancer.

Lambert et al. have previously developed methods for estimating cumulative probabilities of death parametrically in a relative survival framework \cite{Lambert2010}. Eloranta et al. extended these methods to enable partitioning of the excess mortality associated with cancer into component parts such as excess cardiovascular mortality and remaining excess cancer mortality\cite{Eloranta2012}. This has been applied to study temporal trends in excess mortality due to diseases of the circulatory system (DCS) among Hodgkin lymphoma (HL) patients\cite{Eloranta2013}. Further extensions of these methods have been applied to study excess DCS incidence in HL patients\cite{Weibull2019}. In the latter paper, the expected incidence rate was modelled using a constant rate model with two time scales (calendar year and age) with twoway interactions. In the mentioned papers, the cumulative probabilities were calculated using numerical integration, in the presence of competing risks. However, considering morbidity as an absorbing state, as in the application on HL and excess DCS incidence, prevents identification of fatal complications, which in turn makes comparisons with previous studies on excess mortality more difficult.

Multi-state models are a generalisation of competing risks models and one way to study patient trajectories\cite{Putter2007,Iacobelli2013}. The most commonly used method to estimate transition rates in a multi-state model is using a semi-parametric approach, such as the Cox model. In a recently published study, Crowther and Lambert\cite{Crowther2017} suggested a framework for transition-specific parametric distributions in multi-state models. In general, parametric survival models have several advantages over semi-parametric models, including the potential of modelling more than one timescale, more efficient incorporation of time-dependent effects, ease of prediction (both in- and out-of-sample), and greater flexibility in quantifying absolute risks. In addition, estimation of uncertainty is easier, and they enable extrapolation of survival over the full life span (although this should always be done with caution). In the previously proposed method\cite{Crowther2017}, a separate parametric model can be fitted to each transition, which allows the use of transition-specific distributions. The transition probabilities are calculated using a simulation approach\cite{Fiocco2008, Gill1990}. Due to the flexibility of the simulation approach, other quantities of interest can be easily predicted, such as the average time spent in each state. 

In survival analysis, there may be more than one plausible timescale on which to quantify an event rate\cite{Oakes1995}. An obvious example is how incidence, prevalence, and mortality due to various cancers vary as functions of attained age and calendar time. Acknowledging this may introduce transition-specific and/or multiple timescales, which are rarely investigated since the computational complexity increases substantially. Work in this area has been sparse, particularly moving to competing risks and the more general multi-state setting. Lee and Fine\cite{Lee2017a} developed cause-specific Cox models in a competing risks setting, allowing cause-specific timescales, where attained age was the underlying timescale for one cause of death and time since diagnosis was assumed for the competing cause of death. Wolkewitz et al.\cite{Wolkewitz2016} developed a two timescale Cox model applied to modelling risk of infection among patients admitted to Intensive Care Units (ICU), with discharge and death as competing risks, using time since entry to the ICU and calendar time as timescales. In a parametric framework, the most substantial contribution to the field is by Iacobelli and Carstensen\cite{Iacobelli2013}, who provided a general approach for estimating multi-state models with multiple timescales, centred on time-splitting and estimating rates within a parametric spline-based modelling framework using Poisson regression. Predictions of transition probabilities were obtained using simulation, and confidence intervals calculated using a parametric bootstrap. 

Motivated by the need to simultaneously investigate excess morbidity and mortality among cancer survivors, this study proposes one solution to several methodological challenges related to the clinical application. Starting from an illness-death model (with cancer diagnosis as the initial starting state) we incorporate estimation of excess hazard rates into the transition model for the illness state. Moreover, in the expected hazard rate model we allow for multiple time scales (calendar time and age) while remaining transitions have one single time scale (time since study entry or time since entry to current state). Generalisation beyond the illness-death model is easily achieved. The approach is illustrated on a study of excess incidence and mortality from diseases of the circulatory system (DCS) among patients treated for HL.

The article is structured as follows. In Section~\ref{msm} we give a brief overview of recently developed methods for multi-state modelling in the cause-specific framework, after which we describe our proposed approach for each transition. Section~\ref{ex} provides an illustrative example on patients diagnosed with HL and excess DCS morbidity and mortality. We conclude with a discussion and suggest possible extensions in Section~\ref{disc}.

\section{Multi-state modelling incorporating estimation of excess hazards}
\label{msm}
We begin by conceptually describing the method proposed by Crowther and Lambert\cite{Crowther2017}. We define a global vector of available covariates, $\bm{X}$, since the same covariates may be used in multiple models. The fitted transition-specific models can be as complex as required (with respect to functional form of covariates, interactions, and time-varying effects) and one can use different parametric models for different transitions (e.g., a flexible parametric for one transition and a Weibull model for another). Once the set of transition models has been chosen, transition probabilities are calculated for a covariate pattern of interest. The details have been extensively described elsewhere\cite{Crowther2017} but in short, point estimates are obtained by simulating a large sample of patient trajectories ($n$) through the fitted multi-state model. Confidence intervals are calculated from repeated simulations ($m$), each drawing from a multivariate normal distribution with the estimated parameter vector as mean and the variance-covariance matrix (of the parameters) as variance, i.e., a parametric bootstrap.


\begin{figure}[h!]
\centering
\hspace{1cm}
\begin{tikzpicture}
	\path (-5,3.2) node[stor] (ts) { Time scales: \\ $t$: Time since cancer \\ $a$: Age at cancer diagnosis \\ $c$: Year of cancer diagnosis \\ $u$: Time of illness};
 	\path (0,4) node[diskret] (hl) {State 1: \\ Alive with cancer};
	\path (7,5) node[osynlig] (dcs) {State 2:};
	\path	(7,4.2) node[streckad] (expdcs) {Expected illness};
	\path	(7,3.2) node[streckad] (excdcs)  {Excess illness};
	\path	(0,0) node[diskret] (dead)  {State 3: \\ Dead before illness};
	\path	(7,0) node[diskret] (dcsdead){State 4: \\ Dead after illness};
	\path (8,5) node[osynlig] (A) {};
	\draw[thick, solid, rounded corners] ($(dcs.north west)+(-0.2,0.05)$) rectangle ($(excdcs.south east)+(0.2,-0.2)$);
	\draw[->,thick, solid] ($(hl.east)+(0.1,0)$) -- node [text width = 2.2cm, midway, above] {$h_{1}^*(t+a,t+c) $} ($(expdcs.west)+(-0.1,0)$);
	\draw[->,thick, dotted] ($(hl.east)+(0.1,0)$) -- node [text width = 0.9cm, midway, below] {$\lambda_{2}(t)$}  ($(excdcs.west)+(-0.1,0)$);
	\draw[->,thick, solid] ($(hl.south)+(0,-0.1)$) -- node [text width = 1cm, midway, right] {$h_3(t)$} ($(dead.north)+(0,0.1)$);	
	\draw[->,thick,solid] ($(excdcs.south)+(0,-0.3)$) -- node [text width = 1.8cm, midway, right] {$h_4(t-u)$}  ($(dcsdead.north)+(0,0.1)$);
\end{tikzpicture}
\caption{Illustration of suggested approach. $h_{1}^*(t+a,t+c)$ refers to the illness incidence rate that is expected in the absence of cancer with attained age ($t+a$) and calendar year ($t+c$) as time scales, $\lambda_{2}(t)$ is the excess incidence of illness with time since cancer diagnosis ($t$) as the time scale, $h_3(t)$ is the mortality rate with time since diagnosis ($t$) as the time scale, and $h_4(t-u)$ is the mortality rate among cancer survivors who experience the illness, with time since illness ($t-u$) as the time scale.}
\label{fig:illustration}
\end{figure}
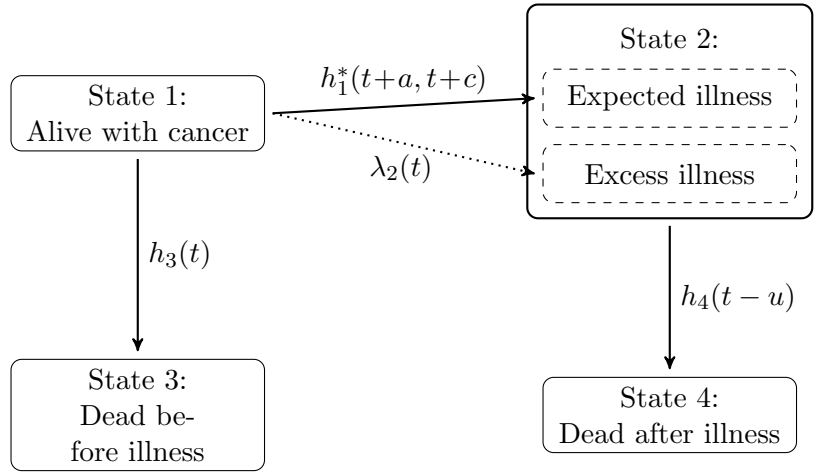

Figure \ref{fig:illustration} shows a representation of our suggested extension. While a standard illness-death model consists of three states (healthy, ill, dead), we partition the illness state into two unobserved states - excess and expected illness (dashed lines). We further have two death states, death following illness, and death without having experienced the illness, which allows us to quantify the impact of the illness under study on mortality. In total, we have four states, of which one is partitioned into two components. Patients start out in state 1; alive after a diagnosis of cancer. Patients might then experience some illness (state 2), which can be either related to the cancer and its treatment (``excess'') or not (``expected''). The transition rate to the expected illness component of state 2, $h_{1}^*(t+a,t+c)$, is modelled using population records of disease incidence with attained age ($t+a$) and calendar time ($t+c$) as time scales. The transition to the excess illness component of state 2, $\lambda_{2}(t)$, is estimated using a parametric excess hazards model, incorporating the modelled expected illness rate. Patients might die prior to experiencing the illness under study (state 3). This mortality rate, $h_3(t)$, can be modelled as all-cause or partitioned into cause-specific mortality models. The time scale for both the excess incidence model and this mortality model is time since cancer ($t$). Some of the patients experiencing the illness (excess or expected) will transition to the absorbing state of death (state 4). This transition, $h_4(t-u)$, can be modelled as an all-cause or cause-specific mortality model, with either one time scale (here we illustrate using time since illness, $t-u$) or multiple time scales (a natural extension would be to add $t$). Any of the models used for transitions into states 3 and 4 can be extended to excess hazards models, given the appropriate expected rate information.

\subsection{Transition 1: Modeling the expected incidence}
In standard relative survival we use mortality rates from publicly available life tables (usually stratified by year, age, and sex) as a measure of the expected mortality in the general population. In order to incorporate such an expected rate into the transition-specific model framework, and hence the simulation approach used to obtain predictions, we must model the rate with an appropriate parametric specification. As incidence, rather than mortality, is the entity of interest here, population incidence tables are required to model the expected incidence rate, i.e., the number of events and person-time for each combination of the stratification variables, as shown in Table~\ref{tab:data_setup}, are needed.

\begin{table}[h!]
\begin{verbatim}
year	  sex      age       d         y
2001     1       18      62   74115.36
2001     1       19      83   74258.12
2001     1       20      85   75230.58
...
2002     1       18      63   70099.6
2002     1       19      64   73899.11
...
\end{verbatim}
\caption{Example of data setup for modelling the expected incidence rate of illness. For every combination of calendar year, sex, and age, the number of events (\texttt{d}) and person-time (\texttt{y}) are calculated.}
\label{tab:data_setup}
\end{table}

While any parametric hazards model can be used to estimate the (log) expected rate, we illustrate how to model the transition from state 1 to the expected illness component of state 2 using a spline-based Poisson regression model. 
\begin{equation}\label{eq:expmodel}
	\log \left( \frac{d}{y}\right) = f(a) + g(c) + \mathbf{x}_{1} \mathbf{\beta}_{1}
\end{equation}

As disease incidence in a population is typically a function of both age and calendar year, both are used as time scales here. Additionally, population incidence/mortality tables are typically not available on individual level, which precludes modelling the rate with one of these variables as the time scale and other as a time-fixed covariate. Age and calendar year can be modelled using a range of functions $f(\cdot)$ and $g(\cdot)$, for example, assuming linearity or using restricted cubic splines or B-splines. In this paper, we use restricted cubic splines to model both timescales, which are easily conducted within a Poisson framework\cite{Iacobelli2013}. 

We will use $\mathbf{x}_{1}$, where $\mathbf{x}_{1} \subset \mathbf{X}$, to denote the vector that contains additional stratification factors for the population incidence table; this typically only contains sex but might also include race, region, or social class. We further let $\mathbf{\beta}_{1}$ denote the vector of associated coefficients. For simplicity, we assume proportional hazards, although this assumption can be relaxed by including interaction terms between the time scales and covariates into the model.

Equation \ref{eq:expmodel} specifies the model for the log expected rate as a function of the two continuous time scales and any additional covariates. However, for implementation purposes, the model needs to be specified in terms of the main time scale for the multi-state model (time since cancer diagnosis): 
\begin{equation}\label{eq:expmodel2}
	\log h^*_{1}(t + a , t + c | \mathbf{x}_{1}) = f(t + a) + g(t + c) + \mathbf{x}_{1} \mathbf{\beta}_{1}
\end{equation}

We provide more details on the implementation in the Appendix.

\subsection{Transition 2: Modeling the excess incidence}
While the expected incidence rate is modelled using population incidence data on a grouped level, the remaining models use the patient cohort data. The setup can be either in wide format (each patient has one row with information on all transitions) or long format as described by Putter et al.\cite{Putter2007} (each patient has a separate row of data for every transition). Long format (i.e., stacked data notation) can be useful in situations where covariate effects are shared across several transitions. We model the excess incidence of illness using a parametric relative survival model. The most appropriate choice of time scale for the excess incidence is time since cancer diagnosis ($t$). Using the notation from Figure~\ref{fig:illustration}, the overall incidence, $h(t,t+a,t+c)$, among the patients can be written as the sum of the incidence expected in the absence of cancer, $h^*_1(t+a,t+c)$, and the excess incidence, $\lambda_2(t)$, associated with the cancer and its treatment:
\begin{equation}\label{eq:relsurvhazard}
	h(t,t+a,t+c | \mathbf{x}_{1}, \mathbf{x}_{2}) = h^*_{1}(t+a,t+c | \mathbf{x}_{1}) + \lambda_{2}(t | \mathbf{x}_{2})
\end{equation}

The vector $\mathbf{x}_{2}$, where $\mathbf{x}_{2} \subset \mathbf{X}$, represents the set of covariates related to the excess rate, which usually includes sex, baseline age and calendar year. We choose to model the transition from state 1 to the excess illness component of state 2 on the log cumulative excess hazard scale. The overall cumulative hazard, $H(t,t+a,t+c | \mathbf{x}_{1},\mathbf{x}_{2})$, can be expressed as the sum of the cumulative expected hazard, $H^*_{1}(t+a,t+c | \mathbf{x}_{1})$, and the cumulative excess hazard, $\Lambda_{2}(t | \mathbf{x}_{2})$:
\begin{equation}\label{eq:relsurvcumH}
	H(t,t+a,t+c | \mathbf{x}_{1},\mathbf{x}_{2}) = H^*_{1}(t+a,t+c | \mathbf{x}_{1}) + \Lambda_2(t | \mathbf{x}_{2}) 
\end{equation} 

Although any parametric relative survival model can be used, we will illustrate using a flexible parametric relative survival model\cite{Nelson2007}, for which the cumulative excess hazard is defined as:
\begin{equation}
	\log \Lambda_2(t | \mathbf{x}_{2}) = \log \Lambda_{2,0}(t) + \mathbf{x}_{2} \mathbf{\beta}_{2}.
\end{equation}

For illustrative purposes, we choose to model with only time since cancer diagnosis as the time scale and we incorporate age and year of diagnosis as baseline covariates (included in $\mathbf{x}_{2}$), but it is possible to incorporate these as secondary time scales. Moreover, interactions, including time-dependent effects, can easily be included. Given the completeness of incident cases of the illness under consideration, the predicted expected rate is treated as fixed, i.e., assuming no uncertainty, tailored at the individual patient level and incorporated into the likelihoood \cite{Brenner2005}.

\subsection{Transition 3: Modeling the mortality rate (before illness)}
In addition to being at risk for late complications, patients are at risk of dying from the cancer itself and from other causes unrelated to the cancer. In the simplest case, interest is not in the mortality from these causes (other than to account for this as a competing event) and any standard parametric all-cause mortality model can be used to model the transition from state 1 to state 3. Note however, that this will implicitly not include deaths caused by the illness under study. We illustrate using a flexible parametric survival model\cite{Royston2002}, where modelling is done on the cumulative hazard scale:
\begin{equation}
	\log H_3(t | \mathbf{x}_{3}) = \log H_{3,0}(t) + \mathbf{x}_{3}\mathbf{\beta}_{3}
\end{equation} 

The log cumulative baseline hazard, $\log H_{3,0}(t)$, is modelled using splines, much like in the excess incidence model above, and we have our covariate vector $\mathbf{x}_{3} \subset \mathbf{X}$ with associated coefficients, $\mathbf{\beta}_{3}$. Although mortality due to other causes would be best modelled using attained age as the time scale, the mortality among these patients is highly driven by cancer deaths, and thus time since cancer diagnosis is the more reasonable choice of time scale. However, we might also extend this model to incorporate multiple time scales. Though not described in detail here, extending this to a more complex model - such as an excess mortality model - is possible. This would introduce also a model for the expected mortality from all causes of death except for the illness.

\subsection{Transition 4: Modeling the mortality rate (after illness)}
Once patients enter the illness state, it is not possible to separate between patients who experienced the illness due to the cancer and patients who would have been diagnosed with the illness regardless of the cancer. So even though there are two transitions going into state 2, i.e., into the expected and excess illness components of the illness state, there is just one transition going from state 2 to whichever states patients can experience next. As with the other transitions, any parametric model can be applied. The choice of time scale can be either time since cancer diagnosis (resulting in a delayed entry model) or time since illness diagnosis, i.e. a clock reset model. It is also possible to use both as time scales. We again illustrate with a flexible parametric survival model, now with time since illness as the time scale. 
\begin{equation}
	\log H_4(t-u | \mathbf{x}_{4}) = \log H_{4,0}(t-u) + \mathbf{x}_{4}\mathbf{\beta}_{4}
\end{equation} 

The log cumulative baseline hazard, $\log H_{4,0}(t-u)$, is modelled using splines, and we have our covariate vector $\mathbf{x}_{4} \subset \mathbf{X}$ with associated coefficients, $\mathbf{\beta}_{4}$. Similar to the transition to death before illness, a more complex model can be incorporated. When applying an excess rate model, the excess mortality could be further partitioned into excess illness- and cancer mortality following the method suggested by Eloranta et al. (2012)\cite{Eloranta2012}. However, this can become complex as it requires calculating expected mortality rates conditioning on the illness event.

For implementation purposes, the excess and expected components that make up state 2 must be treated as separate states (say 2a and 2b) when defining the transition matrix, since they have separate transition hazards governing their rate of occurrence. Subsequently, the transition rates for the next transitions of going from 2a to 3, and 2b to 3, must share the same transition model, as the partitioning of state 2 is essentially theoretical. As a result of this, the accompanying software package has been extended to allow states to share transition model.

\section{Illustrative example}\label{ex}
We illustrate the suggested approach with an application to the modelling of excess incidence and mortality from diseases of the circulatory system (DCS) among patients treated for HL. HL is often pointed out as a model disease in terms of improvements in survival, going from a fatal disease in the mid-1900 to around 90\% of young HL patients today being cured from their lymphoma\cite{Sjoberg2012,Glimelius2015a}. These improvements are primarily due to major advances in the management of these patients, including more accurate radiotherapy, effective multi-agent chemotherapy, and improved staging procedures. With more patients surviving their HL, research focus has somewhat shifted towards understanding and reducing treatment toxicity. The two major complications with potential fatal outcome are secondary malignancies and DCS.

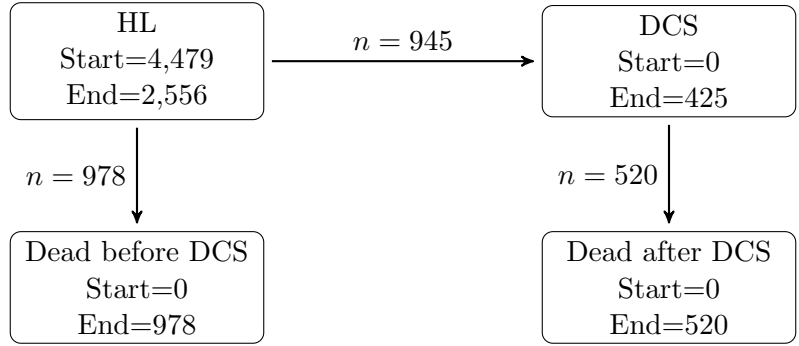
\begin{figure}[h!]
\centering
	\begin{tikzpicture}[node distance=1cm, auto,]
		\path (0,4) node[diskret] (hl) {HL \\ Start=4,479 \\ End=2,556};
		\path (7,4) node[diskret] (dcs) {DCS \\ Start=0 \\ End=425};
		\path (0,1) node [diskret] (dead) {Dead before DCS \\ Start=0 \\ End=978};
		\path (7,1) node[diskret] (dcsdead) {Dead after DCS \\ Start=0 \\ End=520};
		\draw[->,thick,solid] ($(hl.east)+(0.1,0)$) -- node [midway, above] {$n=945$} ($(dcs.west)+(-0.1,0)$);
		\draw[->,thick,solid] ($(hl.south)+(0,-0.1)$) -- node [midway, left] {$n=978$} ($(dead.north)+(0,0.1)$);
		\draw[->,thick,solid] ($(dcs.south)+(0,-0.1)$) -- node [midway, left] {$n=520$} ($(dcsdead.north)+(0,0.1)$);
			\end{tikzpicture}
\caption{Illustration of patient trajectories for the 4,479 patients diagnosed with HL in Sweden between 1985 and 2013, ages 18-80 years at diagnosis. `Start' represents the number of patients who begin in each state, and `End' represents the number of patients remaining in each state at the end of follow-up.}
\label{sec3:states_fig}
\end{figure}

From the Swedish Cancer Register, we identified 4,479 patients diagnosed with HL between 1985 and 2013, aged 18-80 at diagnosis. This cohort has been described in detail previously\cite{Weibull2019}. Patients were followed until death, emigration, or end of study period ($31^{\text{st}}$ of December, 2014), whichever came first. Incident cases of DCS were linked to the HL patients using the Swedish Inpatient Register. Only events with DCS as the main diagnosis were considered. At the end of each individual's follow-up, 2,556 patients had not experienced any of the secondary events, 978 patients died due to any cause prior to experiencing DCS, and 945 patients had been diagnosed with DCS. Among these, 425 were alive and living with DCS at the end of study, and 520 patients had died (from any cause of death) (Figure~\ref{sec3:states_fig}).

For DCS, no population incidence table with sufficiently detailed data is publicly available in Sweden. Instead, a cohort of approximately 10 million individuals (Swedish residents who were either born after 1931 and residing in Sweden after 1961, and their parents, or who were included in a census 1960–1990) was constructed using several population-based registers, as a representation of the Swedish population. Data were stratified by calendar year, age, and sex, and the number of DCS events and amount of person-years at risk were calculated. The expected DCS rate was modelled using Equation \ref{eq:expmodel2}, with 5 degrees of freedom for each time scale and incorporating sex.

For modelling all remaining transitions, flexible parametric survival models were used with year of diagnosis, age at diagnosis, and sex included as baseline covariates. Year and age were parameterised using restricted cubic splines with 5 degrees of freedom in the models for excess DCS incidence and mortality before DCS. For post-DCS mortality, 3 degrees of freedom were used due sparseness of data. No parameter effects were shared between models. For simplicity, only main effects of covariates were considered, however, interaction terms could easily be incorporated. 

The transition from HL to excess DCS was modelled with time since HL diagnosis as the time scale. The predicted expected rate was incorporated into the likelihood to enable estimation of the excess rate. For the transition from HL to dead before DCS, time since HL diagnosis was used as time scale and all-cause mortality was the outcome of interest. Patients with DCS identified from their cause of death certificate ($n=100$) were given that date as their diagnosis date and the following day as their death date. Because of this, the all-cause mortality before DCS implicitly did not include any deaths caused by DCS. Death following a DCS diagnosis was modelled with time since DCS as the underlying time scale (clock-reset).

\begin{figure}[h!]
\centering
	\includegraphics[scale=1.2]{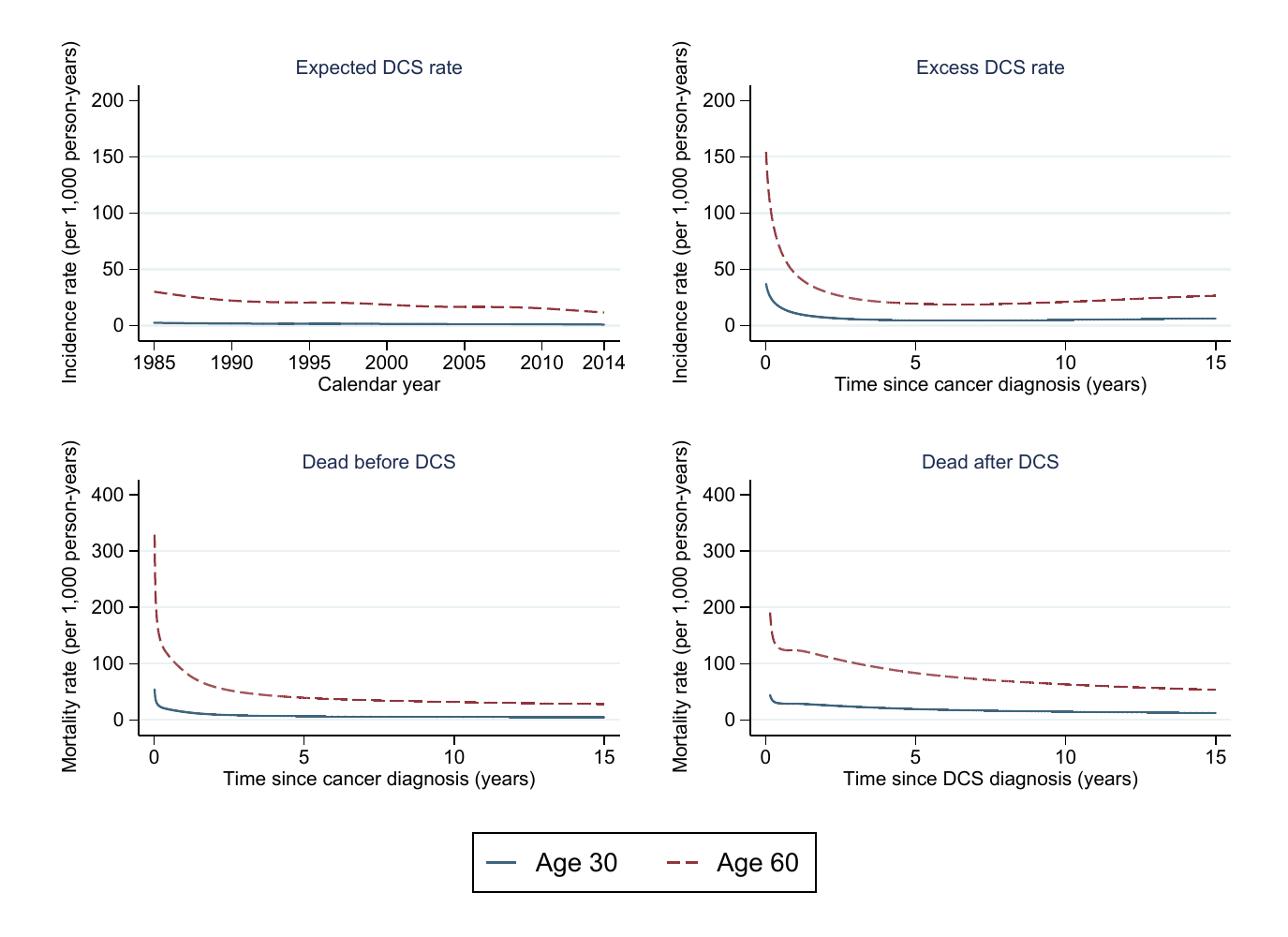}
\caption{Estimated transition rates for male patients aged 30 and 60 years at diagnosis. Expected DCS incidence was modelled as a function of calendar time. Remaining panels are for patients diagnosed in 1995, with excess DCS incidence and other-cause (excluding DCS) mortality before DCS as functions of time since cancer diagnosis, and all-cause mortality after DCS as a function of time since DCS diagnosis.}
\label{sec3:rate_fig}
\end{figure} 

Figure~\ref{sec3:rate_fig} shows the transition rates for male patients aged 30 and 60 year old at diagnosis. The expected DCS rate is presented as a function of calendar time, while the remaining rates are for patients diagnosed in 1995, and plotted over time since diagnosis (cancer or DCS). The expected rates declined slightly following the mid-1980s, and the rate among younger patients was very low throughout. The excess DCS rate and both mortality rates were all high early during follow-up.

\begin{figure}[h!]
\centering
\includegraphics[scale=1.2]{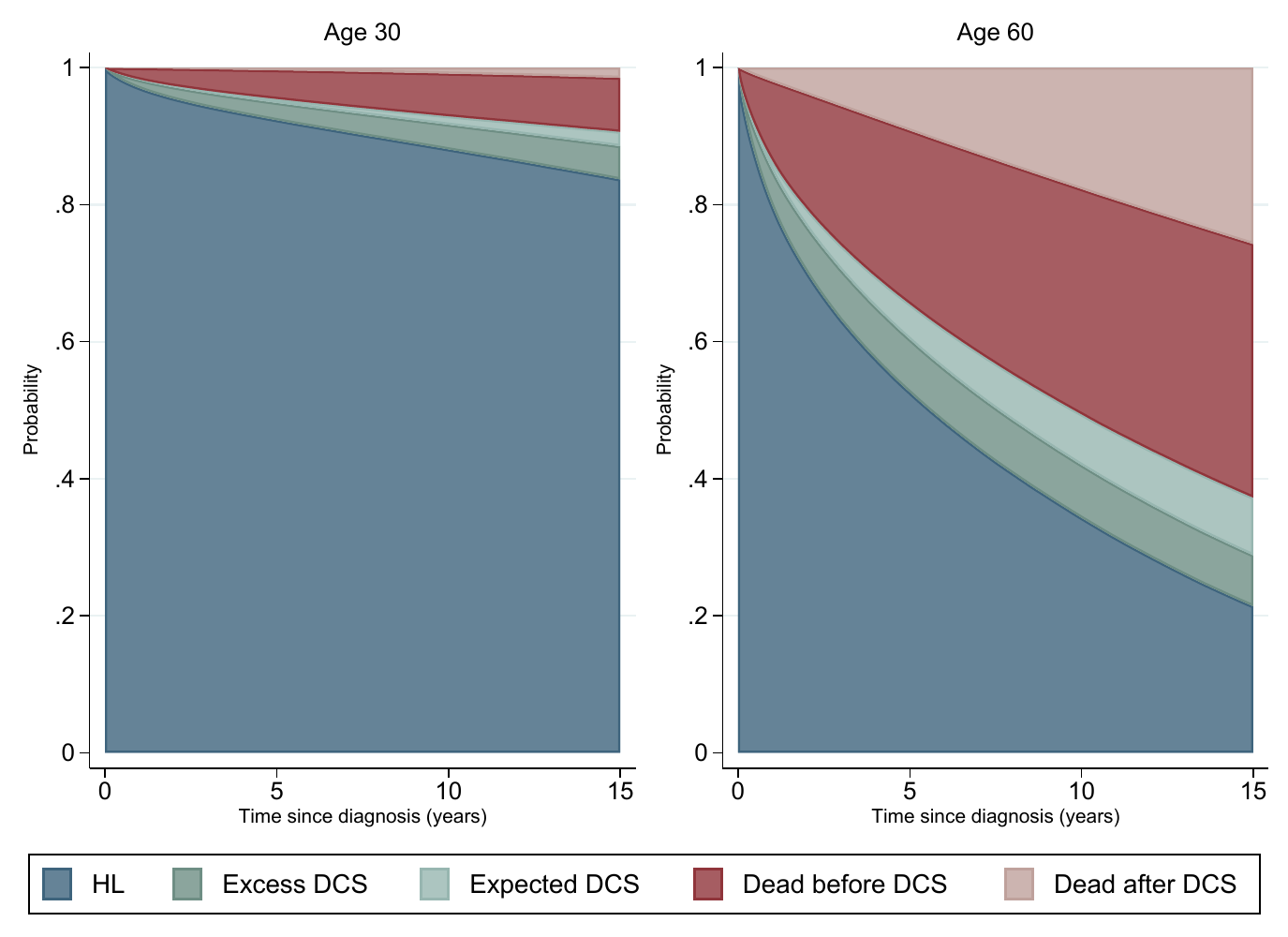}
\caption{Probability of being in each state for male patients diagnosed in 1995, aged 30 and 60 years, as a function of time since HL diagnosis.}
\label{sec3:stacked_fig}
\end{figure}

Using the transition-specific model estimates, point estimate of the transition probabilities were calculated by simulating 1,000,000 subjects through the multi-state model. Figure~\ref{sec3:stacked_fig} shows the stacked transition probabilities over time since HL diagnosis for male patients diagnosed in 1995 at ages 30 (left panel) and 60 (right panel) years at diagnosis. For 30-year-olds, the probability of remaining in the HL state at 15 years after cancer diagnosis was approximately 80\% whereas for 60-year-olds this probability was just above 20\%. The probability of being diagnosed with DCS within the first 15 years as expected in the absence of HL and its treatment, and still being alive, was very small for younger patients. The probability of death (before or after a DCS diagnosis) was expectedly much larger for older patients.

\begin{figure}[h!]
\centering
	\text{(a) Age 30 years at HL diagnosis}\\
	\vspace{0.2cm}	
	\includegraphics[width=0.8\textwidth]{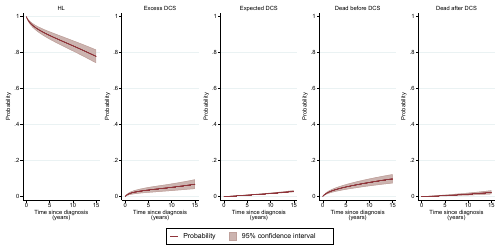} \\
	\vspace{0.1cm}
	\text{(b) Age 60 years at HL diagnosis}\\
	\vspace{0.2cm}	
	\includegraphics[width=0.8\textwidth]{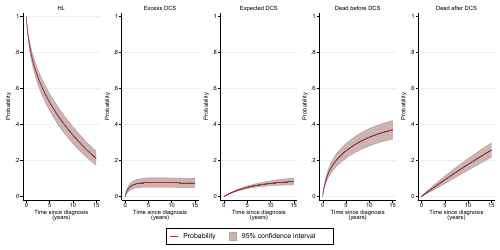}
\caption{Probability (with 95\% confidence interval) of being in each state for male patients diagnosed in 1995, aged 30 (a) and 60 (b) years, as a function of time since diagnosis.}
\label{sec3:ci_fig}
\end{figure}

Although stacked graphs provide a useful tool for illustrating in which state the patients are at different points in time, they lack information on uncertainty. As no parameters were shared between transition models, re-sampling to obtain confidence intervals was done using the transition-specific coefficient vectors and associated variance-covariance matrices, with $n=10,000$ and $m=1,000$. Figure~\ref{sec3:ci_fig} shows the transition probabilities with associated confidence intervals for HL patients diagnosed in 1995 at ages 30 and 60 years. 

\begin{figure}[h!]
\centering
	\text{(a) Age 30 years at HL diagnosis}\\
	\vspace{0.2cm}	
	\includegraphics[width=0.8\textwidth]{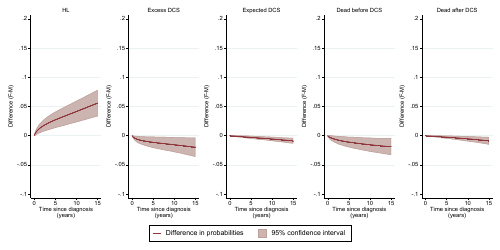} \\
	\vspace{0.1cm}
	\text{(b) Age 60 years at HL diagnosis}\\
	\vspace{0.2cm}	
	\includegraphics[width=0.8\textwidth]{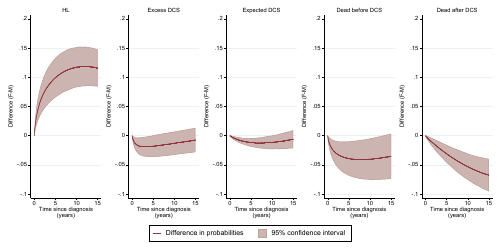}	
\caption{Difference in probabilities (with 95\% confidence interval) of being in each state between female and male patients diagnosed in 1995, aged 30 (a) and 60 (b) years, as a function of time since diagnosis. \textit{Abbreviations; M, males; F, females.}}
	\label{sec3:fig_sex_diffs}
\end{figure}

The fitted models can further be used to predict other clinically relevant measures, such as differences in transition probabilities across varying covariate patterns. Figure~\ref{sec3:fig_sex_diffs} illustrates the differences in transition probabilities between females and males diagnosed in 1995, aged 30 (top panel) and 60 (bottom panel) years. 

\begin{figure}[!h]
\centering
	\includegraphics[width=0.9\textwidth]{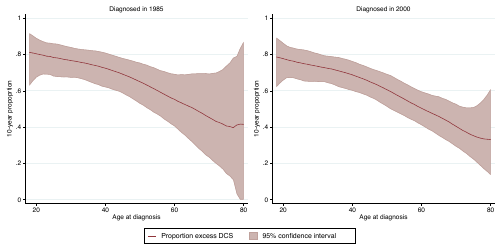}
	\caption{Proportion excess DCS among total DCS probability at 10 years after diagnosis, among male patients diagnosed in 1985 and 2000, by age at diagnosis.}
	\label{sec3:fig_prop_overage}
\end{figure}

Another measure of clinical importance is the proportion of excess DCS (calculated as the probability of excess DCS over the probability of any DCS), which allows us to identify trends in the contribution of excess DCS on total DCS risk as a whole. Figure~\ref{sec3:fig_prop_overage} shows the proportion of excess DCS at ten years after HL, for male patients diagnosed in 1985 and 2000, as a function of age at diagnosis. We could alternatively quantify such a proportion over the other timescales, i.e. as a function of time since diagnosis, or indeed year at diagnosis.

Example code to conduct the previous analysis is included in the Appendix. Extended predictions such as length of stay in each state, the probability of ever visiting a state, standardised predictions and further contrasts are also provided.

\section{Discussion}
\label{disc}
Research on late effects of treatment among cancer survivors is becoming increasingly important as more patients are cured from their disease. Studying life span trajectories requires complex methods, that consider follow-up not just until the first event of interest, but through a series of intermediate states until reaching an absorbing state. Advanced multi-state methods for the cause-specific setting have been developed and are frequently used in applied research, but are unable to separate between the probability of entering an intermediate state associated with a diagnosis of cancer, and the probability of entering the same state not related to cancer. Utilising recently developed methods in multi-state modelling, where a simulation approach is taken together with parametric transition models, we have proposed one way to incorporate estimation of excess morbidity and mortality. This allows for a separation between the predicted excess illness probability and expected illness probability. We have further introduced the possibility for different transitions to have multiple time scales within a relative survival setting.

The developments provided in this paper were motivated by the study of excess morbidity and mortality of DCS among HL patients. The excess DCS mortality has been shown to decline in Sweden since the mid-1970s\cite{Eloranta2013}, and is no longer a major cause of death among this group of patients. Still, the risk of being diagnosed with excess DCS persists in the new millennium among HL patients in Sweden\cite{Weibull2019}. Although a reduction in excess DCS mortality does not necessarily imply a reduction in excess incidence, it remains difficult to draw conclusions on the relationship between HL and its treatment and risk of DCS together with fatality of DCS. By incorporating morbidity and mortality in the same setting, and separating between the incidence attributable to the cancer and that expected in the absence of cancer, we have illustrated one way to overcome this issue. Our approach has been implemented in the \texttt{merlin} and \verb+multistate+ packages in Stata, and code illustrating how to produce the results presented in this paper, can be found in the Appendix. 

For illustrative purposes, we chose to simplify the models in our example. Only main effects models were considered, and no information on clinical covariates were included. Therefore, caution should be taken in terms of interpreting the results presented herein. It is important to point out that more complex, and realistic, models could be fitted to each of the transitions. This includes, but is not limited to, interactions between covariates and relaxing the proportional hazards assumptions \cite{Maringe2019}. As with any standard parametric modelling process, diagnostics and goodness-of-fit inference based on Wald tests, likelihood ratio tests, information criteria (AIC, BIC) can be applied and used to guide model selection for each of the transitions. Lee et al. (2017)\cite{Lee2017a} recently highlighted that standard Martingales must be adapted for the cause-specific timescale setting, and as such, further research is required to adapt them to our more general multiple timescale setting. Indeed, the example model specified for the excess incidence rate (Equation \ref{eq:relsurvhazard}) here only depended on baseline age and calendar year, whereas multiple timescales could be incorporated also for this model. Moreover, the final predicted probabilities rely on four different models so in all applications it is important to assess the goodness of fit of each model. We exemplify this in the Appendix by varying the degrees of freedom; 3-5 degrees of freedom for transition 2 and 3, and 3-4 degrees of freedom for transition 4 (Supplementary Figures ~\ref{app:sensitivity30} and ~\ref{app:sensitivity60}). In our illustrative example we partitioned the illness state into an expected and excess component and focused on one of several possible late effects. However, partitioning the absorbing states can be done in a similar manner given the appropriate expected mortality table, and additional illness states, e.g., secondary malignancies can be added to the model.  

In terms of quantifying model results, we have given examples of how to predict transition probabilities, difference in probabilities, and proportion of excess illness over total probability of illness. Our approach can be used to predict other quantities, such as length of stay in each state (including differences and ratios thereof between patients with different covariate patterns). All of the described extensions can be implemented with the associated software package. 

Several assumptions are made when using a relative survival framework. Patients are assumed to be exchangeable with the general population, conditioned on the stratification variables. If this assumption is not fulfilled, population incidence or mortality tables require additional stratification. If this is not possible, information on a control population can be used to adjust the expected rate \cite{Bower2018}. Additionally, if we wish to interpret the excess rate as treatment-related, then we must assume that the cancer is not directly associated with illness. 

\subsection{Alternative methods}

As in most research applications, there exist multiple routes to achieve the same goal – this is also the case here. An alternative to our chosen approach would have been to design a matched cohort study (i.e., cancer patients would be considered exposed and matched to healthy (unexposed) comparators, on age, sex, and year). This is feasible in countries where appropriate control individual population data is available, however, this is more often not the case, and hence reference population incidence/mortality tables must be utilised. As our focus is on quantifying excess (additive) risk, if a matched cohort approach is taken, additive models would need to be fitted, with the natural model choice being the Aalen additive model \cite{Aalen1989}. Although a simplistic version of this may have been possible, that modelling approach has substantial limitations which we believe strongly precludes its use. We are strong advocates of flexible, yet parametric, modelling approaches. There is limited research into the development of, for example, a spline-based Aalen model, and even more so when attempting to incorporate (arguably essential) time-dependent effects. Taking such an approach would require modelling a baseline rate with additive multiple timescales (which, we would argue, should be modelled in a multiplicative context), combined with an additional additive linear predictor, would raise further challenges. As such, this led to the proposed approach, where we model the baseline (expected) rate appropriately, and combine it with an additive excess component by adopting a relative survival approach. Adopting a flexible parametric approach is important to then obtain smooth estimates of interpretable measures of risk from a comprehensive multi-state model. Having said that, developing an analysis approach for a matched cohort design with multiplicative multiple timescales, and an additive excess component is an important topic for further research, which we are now actively developing as a follow-on study to this paper.

\subsection{Limitations and future research}

In our example, we only considered patients' first occurrence of DCS. In reality, patients may experience multiple DCS occurrences prior to death, or indeed other intermediate events of interest. Incorporating recurrent DCS events raises further challenges, for example, a first occurrence of a DCS event could be cancer-related, whereas a second is not. By introducing subsequent intermediate events into the model framework (with potential further partitioning into excess and expected components), we can accommodate such a setting. This extension is currently being considered.

One could argue that if population level predictions are the entity of interest, marginal predictions rather than covariate-specific, as we have presented, should be derived. Such marginal predictions involve computing the statistic of interest at all observations’ covariate patterns, and taking the average. We have implemented them within the associated software through the \texttt{standardize} option, which is highly computationally intensive in the current implementation, and as such requires further development.

We employed a latent time approach for our simulation method of prediction, also used in Crowther and Lambert (2017)\cite{Crowther2017}; however, Bluhmki er al. (2019)\cite{Bluhmki2019} have highlighted the disconnect between simulating multiple unobservable event times and the real world setting of only observing one outcome. We agree, but would argue that such a hypothetical construct does not preclude the use of the algorithm, and as pointed out by Allignol et al. (2011)\cite{Allignol2011}, quantities obtained from either method are computationally valid. Both methods are now available within the associated software. To derive confidence intervals for all quantities presented in this paper, we have utilized the parametric bootstrap. Alternative techniques, such as the non-parametric bootstrap, may be beneficial as it is robust to model misspecification \cite{Hjort1992}. However, this would be more computationally intensive due to the necessity of re-fitting all transition models on each bootstrap sample. Finally, we may derive standard errors utilising the multivariate delta method to further avoid computationally intensive bootstrapping. Such investigations deserve further research.

\section{Conclusion}

Cancer survivors are at risk of a range of diseases for the remainder of their lives. To support this clinical challenge with scientific real-world evidence, we believe that our approach can be useful for predicting survivors trajectories on a population-based level and to illustrate which of different personal and clinical characteristics are at highest risk of experiencing at different time points following their cancer diagnosis.

\subsection*{Author contributions}
CEW, PCL and MJC contributed to the conception of the work. CEW did the data collection. CEW and MJC implemented the methods, conducted the data analysis and interpreted the findings. CEW, PWD and MJC drafted the article. All authors made critical revision of the article and approved the version to be published. 

\subsection*{Financial disclosure}
None reported.

\subsection*{Conflict of interest}
The authors declare no potential conflict of interests.


\bibliography{manuscript}

\newpage
\appendix

The methods developed in this paper have been implemented in the \texttt{merlin} and \texttt{multistate} Stata packages, which can be installed using:

\begin{itemize}
\item[] \texttt{net install merlin, from(https://www.mjcrowther.co.uk/code/merlin)}
\item[] \texttt{net install multistate, from(https://www.mjcrowther.co.uk/code/multistate)}
\end{itemize}

Further details and examples, on each package, can be found at \texttt{www.mjcrowther.co.uk/software/merlin} and \texttt{www.mjcrowther.co.uk/software/multistate}.

\section*{Example Stata code}
\label{statacode}
Here we detail example syntax that implements the proposed multi-state modelling framework in Stata. Model estimates to obtain the predictions are available upon request.
\begin{stlog}

// Load analysis dataset
use analysisdata, clear

//===========================================================================================================//
// Transition 1: Expected model

// knots for age spline (log scale)
local aknots 2.8904 4.0604 4.2195 4.3175 4.3944 4.5951
// knots for year spline
local yknots 1985 1991 1998 2003 2009 2014

// estimate the spline-based Poisson model and store model object
exptorcs                                              /// function name
                   female,                            /// baseline covariates in expected model
                   expdata(popinc)                    /// dataset with expected rates
                   event(_d) exposure(_y)             /// variables in exprates() file for poisson model
                   year(_year,                        /// year of diagnosis
                             knots(`yknots')               /// knot specification
                             noorthog                      /// do not orthogonalise the spline terms
                   age(_age,                          /// age at diagnosis
                             knots(`aknots')               /// knot specification
                             log                           /// create splines of log age
                             noorthog)                     // do not orthogonalise the spline terms

//store the estimated model
estimates store expmod

//===========================================================================================================//
// We now reshape the analysis data into long format, with one row for each transition at which 
// each patient is at risk for

// Using msset, which will assume a single starting state, and we enter variables 
// which contain event or censoring times for DCS or death. By default, an upper
// triangular transition matrix will be assumed, i.e. the required illness-death model
msset, id(lopnr) states(dcs dead) times(dcs_time dead_time)

// now declare our dataset to be survival data
stset _stop, enter(_start) failure(_status==1) scale(365.24)

//===========================================================================================================//
// Transition 2: Excess model

merlin (_t female rcs(_year, knots(`yknots'))                            ///
                  rcs(_age, knots(`aknots') log)                         ///
                  if _trans==1,                                          ///
                  family(rp, failure(_d) ltruncated(_t0)                 ///
                    df(5) bhazard(exp_dcsrate) noorthog))
estimates store excessmod

//===========================================================================================================//
// Transition 3: Cancer diagnosis to death (no DCS)

merlin (_t female rcs(_year, knots(`yknots'))                            ///
                  rcs(_age, knots(`aknots') log)                         ///
                  if _trans==2,                                          ///
                  family(rp, failure(_d) ltruncated(_t0)                 ///
                    df(5) noorthog))
estimates store deadmod

//===========================================================================================================//
// Transition 4: DCS to death

// We must first generate the clock-reset timescale
gen t2 = _stop - _start

// and re-stset
stset t2, failure(_status==1) scale(365.24)

// estimate the model
merlin (_t female rcs(_year, knots(`yknots'))                            ///
                  rcs(_age, knots(`aknots') log)                         ///
                  if _trans==3,                                          ///
                  family(rp, df(4) noorthog))
estimates store dcsdeadmod

//===========================================================================================================//
// Predictions

// Define transition matrix of the proposed extended illness-death model with the expected and excess 
// components
mat tmat =  (.,1,2,3,.\\.,.,.,.,4\\.,.,.,.,5\\.,.,.,.,.\\.,.,.,.,.)

// Define age and year groups that we want predictions for
local age 30
local year 1995

// Generate a time variable at which to obtain predictions (this will be on the main time since cancer 
// diagnosis timescale)
range tvar 0 15 1000

// Calculate transition probabilities for a male aged 30 at diagnosis, diagnosed in 1995
predictms,   transmatrix(tmat)                 /// 
             models(expmod excessmod deadmod   /// pass model objects for transitions 1 to 5
               dcsdeadmod dcsdeadmod)          ///
             probability                       /// predict transition probabilities      
             timevar(tvar)                     /// time points to predict at
             at1(female 0                      /// specify the covariate pattern to predict at
               _year `year'                    /// specify y, year of diagnosis
               _age `age')                     /// specify a, age of diagnosis
             tsreset(4 5)                      /// transitions 4 and 5 are on the clock-reset timescale
             ci                                 // calculate confidence intervals
	
// Calculate differences in transition probabilities between females and males 
// aged 30 at diagnosis, diagnosed in 1995
predictms,   transmatrix(tmat)                 /// 
             models(expmod excessmod deadmod   /// pass model objects for transitions 1 to 5
               dcsdeadmod dcsdeadmod)          ///
             probability                       /// predict transition probabilities
             timevar(tvar)                     /// time points to predict at
             at1(female 0                      /// specify the 1st covariate pattern to predict at
                _year `year'                    /// specify y, year of diagnosis
               _age `age1')                    /// specify a, age of diagnosis
             at2(female 1                      /// specify the 2nd covariate pattern to predict at
               _year `year'                    /// specify y, year of diagnosis
               _age `age')                     /// specify a, age of diagnosis
             tsreset(4 5)                      /// transitions 4 and 5 are on the clock-reset timescale
             difference                        /// calculate the difference in trans. probs. between at2() - at1()
             ci                                 // calculate confidence intervals

\end{stlog}

\newpage
\section*{Supplementary figures}\label{suppfigs}
\renewcommand{\figurename}{SUPPLEMENTARY FIGURE}
\begin{figure}[h!]
\centering
	\includegraphics[width=0.9\textwidth]{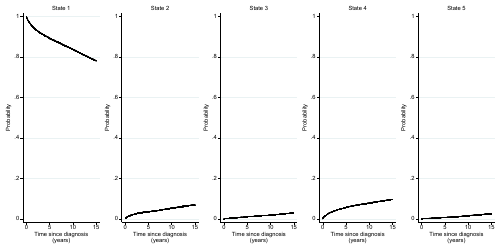}
	\caption{Probability of being in each state for male patients diagnosed in 1995, aged 30 years, as a function of time since diagnosis as predicted from 18 different models with varying degrees of freedom. State definitions: 1 - HL, 2 - Excess DCS, 3 - Expected DCS, 4 - Dead before DCS, 5 - Dead after DCS.}
	\label{app:sensitivity30}
\end{figure}

\begin{figure}[h!]
\centering
	\includegraphics[width=0.9\textwidth]{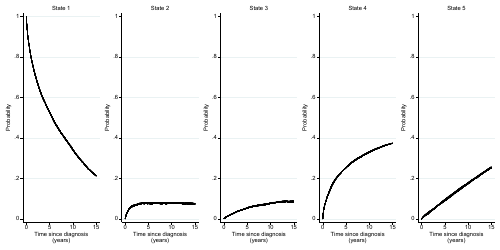}
	\caption{Probability of being in each state for male patients diagnosed in 1995, aged 60 years, as a function of time since diagnosis as predicted from 18 different models with varying degrees of freedom. State definitions: 1 - HL, 2 - Excess DCS, 3 - Expected DCS, 4 - Dead before DCS, 5 - Dead after DCS.}
	\label{app:sensitivity60}
\end{figure}

\end{document}